\documentclass{iau-JDSS}
\usepackage{graphicx}

\title[AGN feedback in numerical simulations] 
{AGN feedback in numerical simulations}

\author[Luca Ciotti]   
{Luca Ciotti}%
\affiliation{Department of Astronomy, University of Bologna, via Ranzani 1, 
40127, Bologna, Italy}

\pubyear{2009}
\volume{Volume 15}  
\pagerange{1}
\date{?? and in revised form ??}
\jname{Highlights of Astronomy, Volume 15}
\editors{Ian F Corbett, ed.}
\begin{document}

\maketitle

\begin{abstract}
  
  The passively evolving stellar population in elliptical galaxies
  (Es) provides a continuous source of fuel for accretion on the
  central supermassive black hole (SMBH), which is 1) extended over
  the entire galaxy life (but declining with cosmic time), 2) linearly
  proportional to the stellar mass of the host spheroid, 3) summing up
  to a total gas mass that is $> 100$ times larger than the currently
  observed SMBH masses, 4) available independently of merging events.
  The main results of numerical simulations of Es with central SMBH,
  in which a physically based implementation of radiative and
  mechanical feedback effects is considered, are presented.

\keywords{X-rays: ISM - Galaxies: cooling flows - Galaxies: active}
\end{abstract}

 
In a series of papers (Ciotti \& Ostriker 2007; Ciotti, Ostriker \&
Proga 2009; Pellegrini, Ciotti \& Ostriker 2009; Shin, Ciotti \&
Ostriker 2009; Jiang \etal\ 2009; see also Ciotti 2009) we study, with
a high-resolution 1-D hydrodynamical code, the evolution of the ISM in
Es under the action of SNIa heating, thermalization of the stellar
mass losses, and feedback from the central SMBH. The cooling and
heating functions include photoionization and Compton effects,
radiation pressure is evaluated by solving the transport equation,
mechanical feedback is produced by a physically based
luminosity-dependent nuclear wind and jet, and star formation is also
allowed. The recycled gas from the aging stars of the galaxy cools and
collapses towards the center, a star-burst occurs and the central SMBH
is fed.  The energy output from the central SMBH pushes matter out,
the accretion rate drops precipitously and the expanding matter drives
shocks into the ISM.  Then the resulting hot bubble ultimately cools
and the consequent infall leads to renewed accretion; the cycle
repeats, with the galaxy being seen alternately as an AGN/starburst
for a small fraction of the time and as a ``normal'' elliptical
hosting an incipient cooling catastrophe for much longer intervals. No
steady flow appears to be possible for Eddington ratios above $\simeq
0.01$: whenever the luminosity is significantly above this limit both
the accretion and the output luminosity are in burst mode.  Strong
intermittencies are expected at early times, while at low redshift the
models are characterized by smooth, very sub-Eddington mass accretion
rates punctuated by rare outbursts.  One of the general consequences
of our exploration is the fact that the recycled gas from dying stars
can induce substantial QSO activity, even in the absence of external
phenomena such as galaxy merging, while accretion feedback can be
strong enough to solve the ``cooling-flow'' problem and to maintain
the mass of the SMBH on the observed range of values.

\end{document}